\documentclass[aps,pra,preprint,superscriptaddress,longbibliography]{revtex4-1}

\usepackage{amsmath}
\usepackage{amsthm}
\usepackage{amsfonts}
\usepackage{amssymb}
\usepackage{xcolor,graphicx}
\usepackage{tikz}
\usepackage{color}
\usepackage[pdftex]{hyperref} 
\usepackage{longtable}
\usepackage{array}
\usepackage{dsfont}

\begin{document}
	
	\title{Critical phenomena in an extended Dicke model}
	
	\author{J. P. J. Rodriguez}
	\email[e-mail: ]{jrod@inaoep.mx}
	\affiliation{Instituto Nacional de Astrof\'{\i}sica, \'Optica y Electr\'onica, Calle Luis Enrique Erro No. 1, Sta. Ma. Tonantzintla, Pue. CP 72840, M\'exico}
	
	\author{S. A. Chilingaryan}
	\affiliation{Instituto Nacional de Astrof\'{\i}sica, \'Optica y Electr\'onica, Calle Luis Enrique Erro No. 1, Sta. Ma. Tonantzintla, Pue. CP 72840, M\'exico}
	
	\author{B. M. Rodr\'iguez-Lara}
	\email[e-mail: ]{bmlara@itesm.mx, bmlara@inaoep.mx}
	\affiliation{Tecnologico de Monterrey, Escuela de Ingenier\'ia y Ciencias, Ave. Eugenio Garza Sada 2501, Monterrey, N.L., M\'exico, 64849.}
	\affiliation{Instituto Nacional de Astrof\'{\i}sica, \'Optica y Electr\'onica, Calle Luis Enrique Erro No. 1, Sta. Ma. Tonantzintla, Pue. CP 72840, M\'exico}

	\date{\today}
	
	\begin{abstract}
Spectral characterization is a fundamental step in the development of useful quantum technology platforms. 
Here, we study an ensemble of interacting qubits coupled to a single quantized field mode, an extended Dicke model that might be at the heart of Bose-Einstein condensate in a cavity or circuit-QED experiments for large and small ensemble sizes, respectively. 
We present a semi-classical and quantum analysis of the model. 
In the semi-classical regime, we show analytic results that reveal the existence of a third regime, in addition of the two characteristic of the standard Dicke model, characterized by one logarithmic and two jump discontinuities in the derivative of the density of states.
We show that the finite quantum system shows two different types of clustering at the jump discontinuities, signaling a precursor of two excited quantum phase transitions.
These are confirmed using Peres lattices where unexpected order arises around the new precursor.
Interestingly, Peres conjecture regarding the relation between spectral characteristics of the quantum model and the onset of chaos in its semi-classical equivalent is valid in this model as a revival of order in the semi-classical dynamics occurs around the new phase transition.
	\end{abstract}
	
	
	\maketitle
\section{Introduction}
The Dicke model describes an ensemble of non-interacting qubits coupled to a single boson mode \cite{Dicke1954p99,Garraway2011}.  It predicts a zero-temperature transition at a critical coupling parameter where the ground state of the model goes from a so-called normal to superradiant phase in the thermodynamical limit \cite{Hepp1973p360,Wang1973p832}. In finite systems, the so-called ground state quantum phase transition (GSQPT) becomes a continuous cross-over where entanglement arises near the critical coupling \cite{Lambert2004,Bakemeier2012,Bao2015}. Such a transition is very hard to observe in the original proposal of non-interacting, two-level neutral atoms coupled to a single electromagnetic field mode at zero temperature, due to restrictions on the achievable coupling strength with respect to the atomic energy gap. Conveniently, theory and experiments involving a Bose-Einstein condensate (BEC) coupled to a high-finesse optical cavity, plus some external standing-wave driving, provide a highly tunable quantum simulation platform to explore the GSQPT as self-organization of the BEC in the optical lattice created by the cavity and driving fields \cite{Domokos2002p253003,Nagy2008127137,Baumann2010p1301,Keeling2010,Nagy2010130401}. In addition, it is also possible to create a simulation of the open Dicke model, with a wider range of independently tunable parameter regimes, coupling two-hyperfine ground states of a BEC using two cavity-assisted Raman transitions \cite{Dimer2007p013804,Baden2014p020408,Zhiqiang2018}. The Dicke model also presents an excited-state quantum phase transition (ESQPT) related to singularities in the spectrum that translate into a logarithmic-type singularity of the semi-classical \cite{Perez2011p033802, Perez2011p046208,Puebla2013} and quantum \cite{Brandes2013,Bastarrachea2014p012004} density of states. The finite size model spectral characteristics might signal a transition from quasi-integrability to non-integrability caused by the quantum precursors of the phase transition that translates into the onset of chaos in the semi-classical equivalent \cite{Emary2003p044101,Emary2003p066203,Bastarrachea2014p032101}. These results impact the dynamics that can be simulated in the circuit and ion-trap quantum electrodynamics (QED) platforms \cite{Chen2007p055803,Mlynek2014,Mezzacapo2015,Barberena2017,Lamata2017,Aedo2018}

Here, we are interested in an extended Dicke model where the qubits are allowed to interact, 
\begin{eqnarray} \label{DLMG}
H = \omega a^{\dagger} a + \omega_{0} J_{z} +  \frac{\gamma}{\sqrt{N_{q}}} \left(a + a^{\dagger}\right) \left( J_{+}+J_{-}\right) +\frac{\eta}{N_{q}}J_{z}^{2}
\end{eqnarray}
where the total number of qubits in the ensemble is $N_{q}$ and their energy gap is given by $\omega_{0}$. The atomic ensemble is described in the orbital angular momentum representation, $J_{i}$ with $ i = x, y, z$, such that $[J_{a},J_{b}] = i \epsilon_{abc} J_{c}$, in terms of a pseudospin length $j= N_{q} / 2$. The field is taken as a boson mode of frequency $\omega$ described by the annihilation (creation) operators, $a$ ($a^{\dagger}$). The qubit ensemble interaction with the boson mode is given by the parameter $\gamma$ and the qubit-qubit interaction is taken as dipole-dipole with nonlinear coupling strength $\eta$. This model might arise in BEC-cavity realizations \cite{Yuan20177404} as well as circuit- and ion-trap-QED. The semi-classical model shows a transition from Rabi to Josephson dynamics where the field is found to break the symmetry of initial symmetric states \cite{RodriguezLara2011p016225}.
Under the rotating wave approximation, the finite size model shows a precursor of the GSQPT that provides entanglement and its semi-classical analog shows a transition from order to disorder at a critical energy related to the spectral characteristics \cite{Robles2015p033819}. 
In the following, we will conduct a detailed analysis of the full model, paying particular attention to the (ESQPT) which has not been looked at so far.
For this, we will find the semi-classical density of states and compare it with the quantum density of states for an ensemble composed of $200$ qubits. 
Our results show the existence of a new third spectral regime outlined by the existence of one logarithmic and two jump discontinuities in the derivative of the density of states. 
Then, we will look at the Peres lattice for the $z$-component of the quantum angular momentum in the three regimes to find the critical energies that may signal a transition from order to disorder in the semi-classical dynamics.
Interestingly, the semi-classical dynamics in the third regime show a transition from order to disorder and, then, additional islands of order appear.
We close this manuscript with our conclusions.

\section{Semi-classical critical analysis}

\begin{figure}[h!]
	\includegraphics[scale=1]{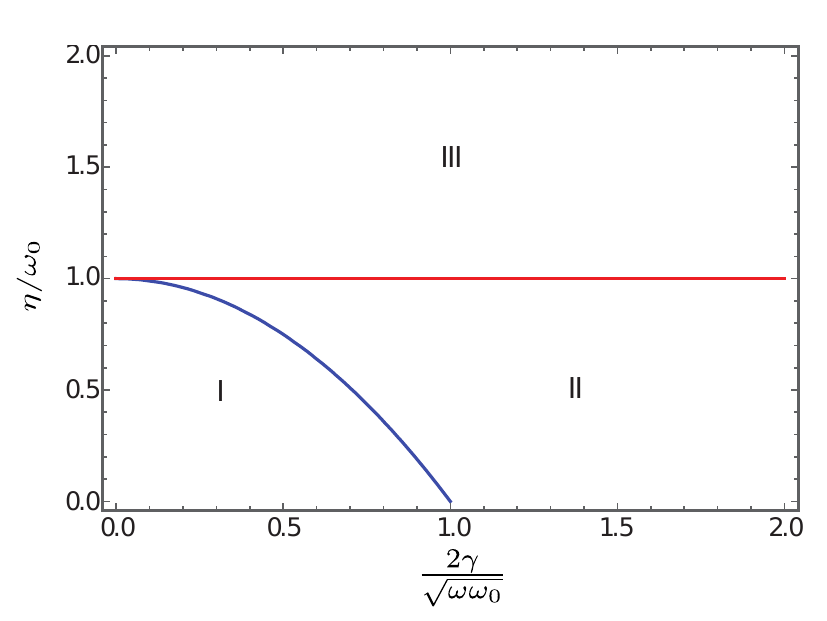}
	\caption{Regions in parameter space classified via types of fixed points of the classical Hamiltonian in Eq.(\ref{eq:2}). Three regions with different fixed point structured are found: Region I,  $ f < 1 $, region II, $ f \geq 1 $ and $ \eta < \omega_{0} $, and region III,  $ f \geq 1 $ and $ \eta \geq \omega_{0} $.} \label{fig:Figure1}
\end{figure}

The extended Dicke model in Eq.(\ref{DLMG}) is not analytically solvable, but we can provide analytical closed-form expressions for the fixed points of the energy surfaces, critical energies, and density of states (DoS) of its semi-classical equivalent. 
These structures will serve as a reference for the numerical analysis of the finite size quantum model.
We obtain the semi-classical Hamiltonian, 
\begin{eqnarray} \label{eq:2}
H_{cl} = \frac{\omega}{2}\left( q^{2} + p^{2} \right) + \omega_{0} j_{z}  + 2 \gamma q \left(\frac{j^{2} - j_{z}^2}{j}\right)^{1/2}   \cos \phi + \frac{\eta}{2 j} j_{z}^2,
\end{eqnarray}
in the usual way \cite{Aguir1992p291}. 
We replace the angular momentum operators with their classical counterparts: $j$ for the total momentum, $j_{z}$ for its projection on the $z$-direction, and $j_{x} = \sqrt{j^2 - j_{z}^2}~ \cos \phi $ for the projection in the $x$-direction. 
For the field we use the classical analogue of the field quadratures,  $\hat{q} = \left(\hat{a}^{\dagger} + \hat{a} \right)/\sqrt{2}$ and $\hat{p} = i \left(\hat{a}^{\dagger} - \hat{a} \right)/\sqrt{2}$.
The semi-classical equations of motion,
\begin{eqnarray}
&& \frac{dq}{dt}  =  \omega p, \\
&& \frac{dp}{dt}  =  - \omega q - 2 \gamma \left(\frac{j^{2} - j_{z}^2}{j}\right)^{1/2}  \cos \phi, \\
&& \frac{d \phi}{d t} = \omega_{0} + \frac{j_{z}}{j} \left[ \eta - 2 \gamma q \left(\frac{j^{2} - j_{z}^2}{j}\right)^{-1/2}  \cos \phi \right], \\
&& \frac{d j_{z}}{dt} = 2 \gamma q \left(\frac{j^{2} - j_{z}^2}{j}\right)^{1/2}   \sin \phi,
\end{eqnarray}
yield six fixed points on the energy surfaces. 
Two fixed points occur for any given set of parameters $\{\eta,\gamma\}$,
\begin{eqnarray}
\left\lbrace  q, p, j_{z} \right\rbrace = \left\lbrace 0, 0, \pm j \right\rbrace, 
\end{eqnarray}
whose nature depends on the auxiliary parameter
\begin{eqnarray}
f = \frac{4 \gamma^2 + \eta \omega}{\omega \omega_{0}}.
\end{eqnarray}
The fixed point $\left\lbrace  q, p, j_{z} \right\rbrace = \left\lbrace 0, 0,  j \right\rbrace$ is not stable for any given value of $f$, while the fixed point $\left\lbrace  q, p, j_{z} \right\rbrace = \left\lbrace 0, 0,  -j \right\rbrace$ is stable for $f<1$ and becomes unstable for $f\geq1$.
In the case when $f\geq1$ and $\eta<w_0$, we find two well-known stable fixed points\cite{Robles2015p033819},
\begin{eqnarray}
\left\lbrace  q, p, j_{z}, \phi  \right\rbrace = \left\lbrace - q_{(s)}, 0, -j f^{-1}, 0 \right\rbrace, \left\lbrace  q_{(s)}, 0, -j f^{-1}, \pi \right\rbrace,
\end{eqnarray}
with $q_{(s)} =2 \gamma \left( j - j f^{-2} \right)^{1/2} / \omega$.
The final two fixed points are obtained for $f\geq1$ and $\eta\geq w_0$,
\begin{eqnarray}
\left\lbrace  q, p, j_{z}, \phi  \right\rbrace = \left\lbrace 0, 0, - \frac{\eta j_{z}}{\omega_{0}} , \pm \frac{\pi}{2} \right\rbrace.
\end{eqnarray}
We can use these six fixed points to divide the parameter space provided by the ensemble-field and qubit-qubit couplings into three regions as shown in Fig.\ref{fig:Figure1}. 
In region I, $f<1$, there are two fixed points, a local maximum and a global minimum. 
The global minimum becomes a saddle point and two degenerate minima emerge in region II for $f\geq1$ and $\eta<w_0$.
In the final region, $f\geq1$ and $\eta\geq w_0$, the saddle point from region II transforms into a local maximum and two degenerate saddle points appear.

\begin{figure}[h!]
	\includegraphics[scale=1]{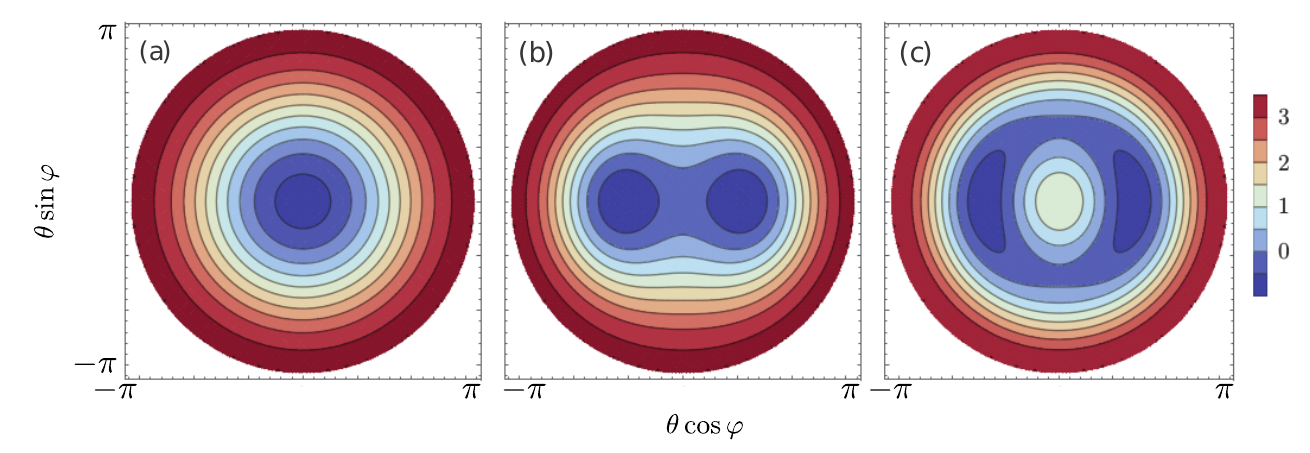}
	\caption{Energy surfaces of the model on resonance, $\omega= \omega_{0}$, in (a) region I, $ f < 1 $ with $\eta = 0.2 ~\omega_{0}$ and $\gamma = 0.3 ~\omega_{0}$ (b) region II, $ f \geq 1 $ and $ \eta < ~\omega_{0}$ with $\eta = 0.2 ~\omega_{0}$ and $\gamma = 0.8 ~\omega_{0}$,  and (c) region III, $ f \geq 1 $ and $ \eta \geq ~\omega_{0} $ with $\eta = 2.1 ~\omega_{0}$ and $\gamma = 0.6 ~\omega_{0}$.  } \label{fig:Figure2}
\end{figure}

The energy minima in these three regions can be obtained by evaluating the Hamiltonian at the stable fixed points,
\begin{eqnarray}
\epsilon_{min} \equiv \frac{E_{min}}{\omega_{0} j} = 
\left\{ \begin{array}{ll}
-1 + \frac{\eta}{2 \omega_{0}} , &  f < 1, \\ 
-\frac{1}{2} \left( f + f^{-1} \right) + \frac{\eta}{2 \omega_{0}} , & f \geq 1,
\end{array} \right.
\end{eqnarray}
Figure \ref{fig:Figure2} shows contour plots of the energy surface for the model in the three regions defined by the fixed points. 
In region I, $f < 1$, there is a local maximum with energy,
\begin{eqnarray}
\epsilon_{+} = 1 + \frac{\eta}{2 \omega_{0}},
\end{eqnarray}
that is above the global minimum, $\epsilon_{min} < \epsilon_{+}$, Fig.\ref{fig:Figure2}(a). In the parameter region II,  $ f \geq 1 $ and $ \eta < \omega_{0} $, there is a saddle point with energy,  
\begin{eqnarray}
\epsilon_{-} = - 1 + \frac{\eta}{2 \omega_{0}},
\end{eqnarray}
and one local maximum with energy $\epsilon_{+}$. 
It is straightforward to check that $ \epsilon_{min} < \epsilon_{-} < \epsilon_{+} $ in this region, Fig.\ref{fig:Figure2}(b).
Finally, in region III, $f \geq 1 $ and $ \eta \geq \omega_{0}$, a saddle point with energy,
\begin{eqnarray}
\epsilon_{s} \equiv - \frac{\omega_{0}}{2 \eta},
\end{eqnarray}
emerges along with two local maxima with energies $\epsilon_{\pm}$. Again, it is straightforward to see that $ \epsilon_{min} < \epsilon_{s} < \epsilon_{-} < \epsilon_{+} $ in this region, Fig.\ref{fig:Figure2}(c).

\begin{figure}[h!]
	\includegraphics[width=\columnwidth]{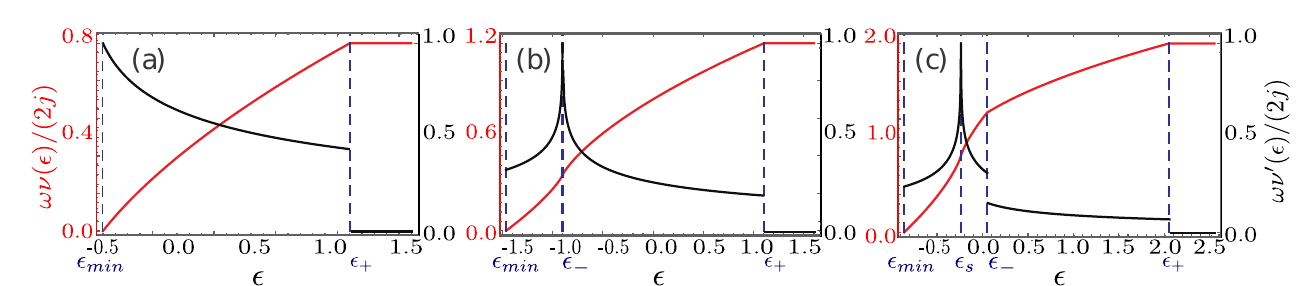}
	\caption{ Scaled semi-classical DoS, $\omega \nu (\epsilon) / (2 j)$ (red), and its first derivative (black) in terms of the scaled energy $\epsilon \equiv E / (\omega_{0} j) $ for parameters identical to those in Fig.\ref{fig:Figure2} } \label{fig:Figure3}
\end{figure}

The critical energies calculated above help us calculating an analytic density of states (DoS) for the classical model in terms of the so-called Weyl's law \cite{Gutzwiller},
\begin{eqnarray}
\nu (E) = \frac{1}{(2 \pi)^2} \int dq ~ dp ~d\phi ~dj_{z} ~ \delta \left( E - H_{cl}\left(q, p, \phi, j_{z}\right) \right), 
\end{eqnarray}
that determines the allowed phase space volume for a given energy $E$. 
The integration over the bosonic canonical pair, $q$ and $p$, is readily performed and yields a constant equal to $ 2 \pi / \omega $. The pseudo spin part of the integral is restricted by the following condition,
\begin{eqnarray}
(1 - z^{2}) \cos^{2} \phi \geq \frac{\omega \omega_{0}}{2 \gamma^{2}} \left( \frac{\eta}{2 \omega_{0}} z^{2} + z - \epsilon \right),
\end{eqnarray}
where we defined the ratio of the $z$-projection to the total orbital angular momentum as the new integration variable, $ z = j_{z} / j $, and a scaled energy, $\epsilon = E / (\omega_{0} j)$. 
In region I, we recover a DoS with two subregions,   
\begin{eqnarray}
\frac{\omega}{2 j}\nu(\epsilon) = 
\left\{ \begin{array}{ll}
\frac{1}{\pi} \int_{z_{2}}^{z_{+}} \phi_{0} (z, \epsilon) dz + 	\frac{z_{2} + 1}{2} +, &  \epsilon_{-} \leq \epsilon \leq \epsilon_{+}, \\ 
1 , & \epsilon_{+} < \epsilon,
\end{array} \right.
\end{eqnarray}
whose derivative shows a discontinuity of the so-called jump-type at the critical energy $\epsilon_{+}$, Fig.\ref{fig:Figure3}(a). 
This might be taken as a semi-classical signature of the ESQPT.
In region II, three different DoS subregions are identified, 
\begin{eqnarray}
\frac{\omega}{2 j}\nu(\epsilon) = 
\left\{ \begin{array}{ll}
\frac{1}{\pi} \int_{z_{-}}^{z_{+}} \phi_{0} (z, \epsilon) dz, &  \epsilon_{min} \leq \epsilon \leq \epsilon_{-}, \\
\frac{1}{\pi} \int_{z_{2}}^{z_{+}} \phi_{0} (z, \epsilon) dz + \frac{z_{2} + 1}{2}, &  \epsilon_{-} < \epsilon \leq \epsilon_{+}, \\ 
1 , & \epsilon_{+} < \epsilon.
\end{array} \right.
\end{eqnarray}
At the critical energy $\epsilon_{-}$, the DoS derivative shows a logarithmic-type discontinuity and the jump-type discontinuity remains at $\epsilon_{+}$,  Fig.\ref{fig:Figure3}(b).  
This behavior is characteristic of the Dicke model and signals the existence of two essentially different ESQPT at energies $\epsilon_{\pm}$ \cite{Bastarrachea2014p012004}.
In region III, $ f \geq 1 $ and $ \eta \geq \omega_{0} $, we find a behavior different from the standard Dicke model. Four different DoS subregions appear, 
\begin{eqnarray}
\frac{\omega}{2 j}\nu(\epsilon) = 
\left\{ \begin{array}{ll}
\frac{1}{\pi} \int_{z_{-}}^{z_{+}} \phi_{0} (z, \epsilon) dz, &  \epsilon_{min} \leq \epsilon \leq \epsilon_{s}, \\
\frac{1}{\pi} \left[ \int_{z_{-}}^{z_{1}} \phi_{0}(z, \epsilon) dz + \int_{z_{2}}^{z_{+}} \phi_{0}(z, \epsilon) dz \right] + \frac{z_{2}-z_{1}}{2}, & \epsilon_{s} < \epsilon \leq \epsilon_{-}, \\ 
\frac{1}{\pi} \int_{z_{2}}^{z_{+}} \phi_{0} (z, \epsilon) dz + 	\frac{z_{2} + 1}{2}, &  \epsilon_{-} < \epsilon \leq \epsilon_{+}, \\ 
1 , & \epsilon_{+} < \epsilon.
\end{array} \right.
\end{eqnarray}
The logarithmic-type discontinuity relocates at the critical energy $\epsilon_{s}$, related to the new saddle points in the energy surface, a new jump-type discontinuity in the DoS derivative appears at $\epsilon_{-}$, and the jump-type discontinuity at $\epsilon_{+}$ remains signaling three possible ESQPT in region III, Fig.\ref{fig:Figure3}(c). 

In all these expressions, the auxiliary parameters $z_{\pm}$, fulfilling $z_{-} \leq z_{+}$, are the real roots of the following quadratic equation,
\begin{eqnarray}
(1 - z^{2}) = \frac{\omega \omega_{0}}{2 \gamma^{2}} \left( \frac{\eta}{2 \omega_{0}} z^{2} + z - \epsilon \right),
\end{eqnarray} 
the parameters $z_{1,2}$, with $z_{1} \leq z_{2}$, are the real roots of the quadratic equation, 
\begin{eqnarray}
\frac{\eta}{2 \omega_{0}} z^{2} + z - \epsilon = 0.
\end{eqnarray} 
and the function $\phi_{0}(z, \epsilon)$ is given by,
\begin{eqnarray}
\phi_{0}(z, \epsilon) = \arccos \left[ \frac{\omega \omega_{0}}{2 \gamma^{2}} \frac{ \left(  \frac{\eta}{2 \omega_{0}} z^{2} + z -\epsilon \right) }{1-z^{2}} \right]^{-1/2}.
\end{eqnarray}

\begin{figure}[h!]
 	\includegraphics[width=\columnwidth]{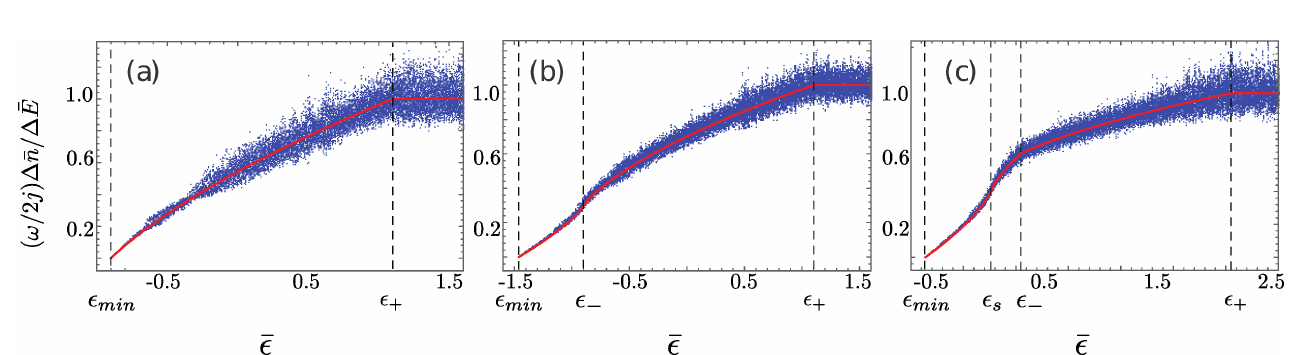}
	\caption{Scaled averaged quantum DoS, $\frac{\omega}{2 j} \bar{\nu}(\bar{\epsilon})$ (blue dots), and its semi-classical analogue (red) as a function of the scaled energy, $\bar{\epsilon}$ for parameters identical to those in Fig.\ref{fig:Figure3}.} \label{fig:Figure4}
\end{figure}

Now, we have well defined semi-classical signatures of possible ESQPTs. 
A ESQPT refers to a singularity in the energy spectrum caused by a change in the clustering of excited states at a critical energy \cite{Capiro2008p1106}. 
Therefore, it is directly manifested in the density of states as discontinuities or divergences \cite{Bastarrachea2014p032101}.  
Unfortunately, the extended Dicke model remains an unsolved model and we are restricted to a numerical analysis of finite, truncated, computational realizations. 
Finite models do not show a sharp quantum phase transitions but smooth crossovers from different spectral configurations.
Nevertheless, the semi-classical results provide a valuable starting point to search for precursors of ESQPTs in the finite extended Dicke model.
In our numerics, we use an ensemble composed of two hundred qubits, $N_{q}=200$, and an extended bosonic coherent basis \cite{Chen2008p051801} with a maximum of six hundred bosons in the field mode, $n_{max}=600$. 
We restrict our analysis to the positive parity subspace of the model and obtain about fifty thousand converged eigenstates, with a wavefunction convergence criteria of less than $10^{-18}$ \cite{Bastarrachea2014p012004}.
This allows us to calculate a numerical averaged quantum DoS,
\begin{eqnarray}
\bar{\nu}(\bar{\epsilon}) = \frac{\Delta \bar{n}}{\Delta \bar{E}},
\end{eqnarray}
as a function of the scaled energy,
\begin{eqnarray}
\bar{\epsilon} = \frac{1}{\omega N_{q}} \left[ \bar{E}(\bar{n}+1) - \bar{E}(\bar{n})\right],
\end{eqnarray}
where the average energy, $\bar{E}(\bar{n})$, and number of photons, $\bar{n}$, are taken over twenty eigenvalues. 
Figure \ref{fig:Figure4} shows the averaged quantum DoS as blue dots with its semi-classical analogue as a solid red line for comparison.
It is possible to see that the averaged quantum DoS follows the trend of its semi-classical equivalent, and shows clustering in the spectrum near the critical scaled energies where the ESQPT is expected.

Peres lattices are an alternative qualitative method to find the precursors of ESQPTs.
Originally conceived as a visual test for the competition between regular and chaotic features in the semi-classical equivalent of quantum models \cite{Peres1984p1711}, the idea behind this method is simple.
If we consider an integrable quantum system described by the Hamiltonian $H_{0}$ and a constant of motion $I$, such that $ \left[ H_{0}, I \right] = 0 $, and plot the mean value of the constant of motion for each and every spectral state versus its energy, we will observe a lattice formed by regularly distributed points because each spectral state can be labeled by the quantum number associated with the observable. 
Introducing a perturbation, $H^{\prime}$, may render the system non-integrable. 
In such a case, the observable $I$ is no longer a constant of motion and the spectrum states cannot be labeled uniquely by a combination of the energy and the mean value of the observable.
However, a weak perturbation might not entirely destroy the regular lattice obtained before but, as the perturbation grows, the regular part of the lattice will disappear gradually and disorder will dominate. 
Thus, the method of Peres lattices serves as an indicator of the changing structures inside the quantum spectrum of the system and has proven a useful method for identifying the various types of ESQPT in the standard Dicke model \cite{Bastarrachea2014p032102}.

\begin{figure}[h!]
	\includegraphics[width=\columnwidth]{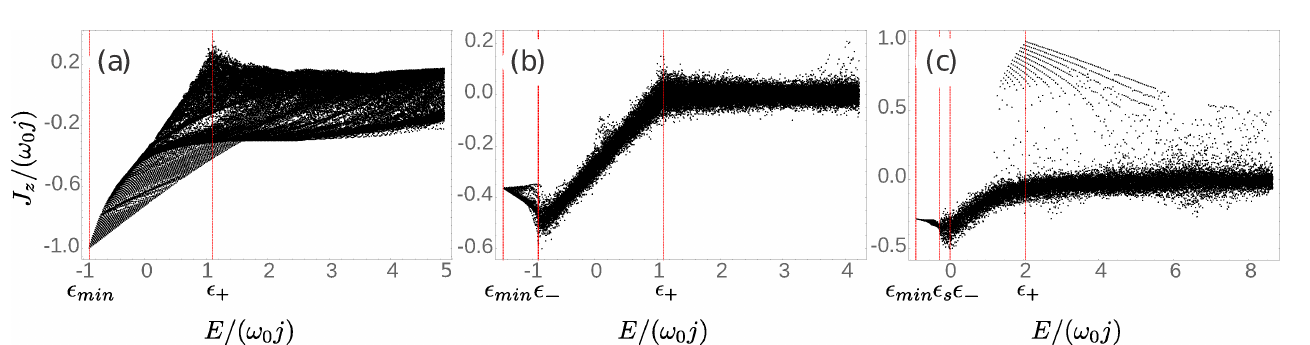}
	\caption{Peres lattice of the scaled quantum angular momentum $ \langle J_{z} \rangle / ( \omega_{0} j ) $ for parameter values identical to those in Fig.\ref{fig:Figure4}.} \label{fig:Figure5}
\end{figure}

\begin{figure}[h!]
	\includegraphics[scale=1]{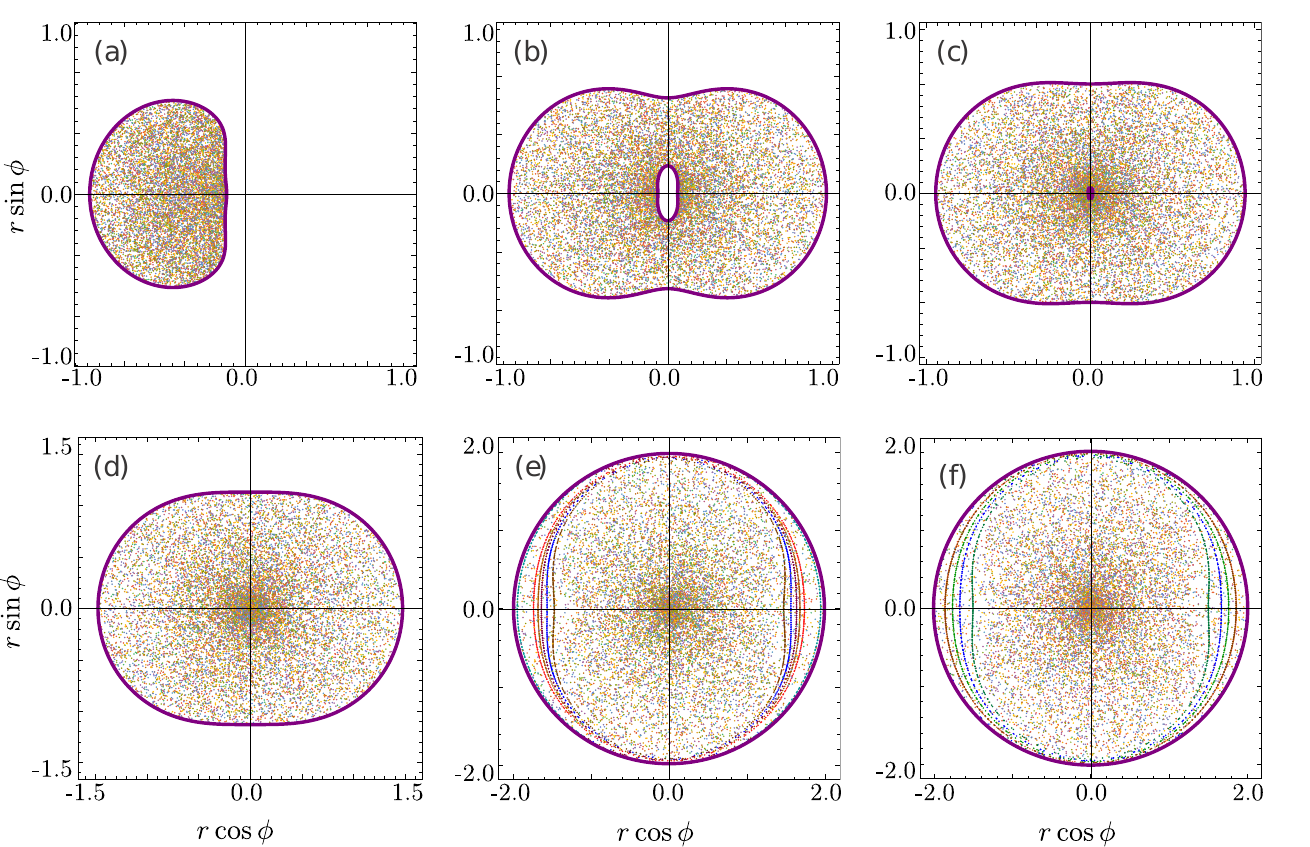}
	\caption{Poincar\'e sections in the semi-classical equivalent of the extended Dicke model for the phase space $ \left\lbrace r, \phi \right\rbrace$ at $ p(t) = 0 $ with $ r = 1 + j_{z} / j $ and initial scaled energies (a) $ E / {(\omega_{0} j)} = -0.3 $, (b) $ E / {(\omega_{0} j)} = -0.15 $, (c) $ E / {(\omega_{0} j)} = 0 $, (d) $ E / {(\omega_{0} j)} = 0.15 $, (e) $ E / {(\omega_{0} j)} = 2 $ and (f) $E / {(\omega_{0} j)} = 2.1 $ The parameters belong to the region III: $\omega = \omega_{0} = 1, $ $\eta = 2.1 $, $ \gamma = 0.6 $, $ j = 100 $.  } \label{fig:Figure6}
\end{figure}

When we look at the Peres lattice for the $z$-projection of the angular momentum operator in the extended Dicke model, Fig.\ref{fig:Figure5}, those for region I and II are phenomenologically identical to those found in the standard Dicke model  \cite{Bastarrachea2014p032102}.
The precursor of the so-called static ESQPT, associated with a maximum of the scaled quantum angular momentum, occurs around the critical scaled energy $\epsilon_{+}$ in region I, Fig.\ref{fig:Figure5}(a), II, Fig.\ref{fig:Figure5}(b), and III, Fig.\ref{fig:Figure5}(c). 
The precursor for the dynamic ESQPT, associated with a minimum of the scaled angular momentum, appears only in region II and III around the critical energy $\epsilon_{-}$.
Region III deviates from the standard Dicke model behavior, here the large values of the nonlinear coupling, restore regularity to the Peres lattice around the critical energy for the precursor of the static ESQPT for large values of the scaled angular momenta, Fig.\ref{fig:Figure5}(c). 
Peres lattice in region III tells us that, for small energies, the available phase space in the semi-classical analogue, provided in terms of the scaled angular momentum projection, will be highly restricted and asymmetric below critical energy value $\epsilon_{s}$, Fig.\ref{fig:Figure6}(a). 
It will become symmetric around $\epsilon_{-}$, Fig.\ref{fig:Figure6}(b), and start expanding, Fig.\ref{fig:Figure6}(c-d), until it reaches its maximum near $\epsilon_{+}$, Fig.\ref{fig:Figure6}(e-f).
As expected from Peres conjecture, the trajectories in the semi-classical analogue will be chaotic for parameter values in the irregular lattice,  Fig.\ref{fig:Figure5}(a-c), and regular for those corresponding to ordered sections of the lattice, Fig.\ref{fig:Figure6}(e-f).
This revival of the regular domain induced by the nonlinear interaction is the most striking effect of the model.

\section{Conclusion}

We studied a Dicke model with dipole-dipole interacting qubits. 
The semi-classical equivalent of the quantum model allowed us to provide a detailed analysis of the energy landscape, where three structurally distinct regions can be identified.
These regions show two, three, and four critical energies at which minima, maxima, and saddle points of the energy manifold appear.
The semi-classical model allowed us to calculate a closed-form density of states that shows a jump-type discontinuity in the first region, a logarithmic- and jump-type in the second region, and a logarithmic- and two jump-type discontinuities in the third region at the critical energies.
Our results served as pointers to focus the search of precursors of ground and excited quantum phase transitions in the quantum model.

We diagonalized the finite-size quantum model using an extended coherent state basis in the positive parity sector of the related Hilbert space.
Our numerical realization considered an ensemble of two hundred qubits with a maximum of six hundred excitations in the boson field and yielded an approximate of fifty thousand converged eigenstates and their respective eigenvalues.
The resulting averaged quantum density of states followed in good agreement the trend provided by the semi-classical analytic result signaling the precursors of quantum phase transitions.
The first two regions yield results phenomenologically identical to those of the Dicke model, with critical energies displaced by the nonlinear coupling.
In the third region, a large nonlinear coupling produces an irregular Peres lattice for the $z$-component of the angular momentum for low energies, differing from the behavior in the standard Dicke model, and a small  regular section arises  for large value of the angular momentum projection near the critical energy related the second jump-type discontinuity in the derivative of the semi-classical density of states. 
The parameters associated with this region produces a revival of regular semi-classical trajectories in an otherwise chaotic system.

We hope that this semi-classical and quantum analysis of the Dicke model with interacting qubits might shed light on the dynamical regimes available for simulations of the model in cavity, ion-trap, and cirquit quantum electrodynamics platforms.

\begin{acknowledgments}
B.M.R.L. acknowledges fruitful discussion with F\'elix Humberto Maldonado Villamizar and Benjam\'in Raziel Jaramillo \'Avila. J.P.J.R acknowledges funding from Consejo Nacional de Ciencia y Tecnolog\'ia (CONACYT) (CB-2015-01-255230) and B.M.R.L from Consejo Nacional de Ciencia y Tecnolog\'ia (CONACYT) (CB-2015-01-255230, FORDECYT-296355).
\end{acknowledgments}

%

\end{document}